\documentclass[prl,twocolumn]{revtex4}

\usepackage{amsmath}
\usepackage[english]{babel}

\begin{document}

\

\title{Comment on ``Nongeometric Conditional Phase Shift via
Adiabatic Evolution of Dark Eigenstates: A New Approach to Quantum
Computation''}

\author{Ognyan Oreshkov and John Calsamiglia}

\affiliation{Grup de F\'{i}sica Te\`{o}rica, Universitat Aut\`{o}noma de
Barcelona, 08193 Bellaterra (Barcelona), Spain}
\date{\today}

\pacs{03.67.Lx, 03.65.Vf}

\maketitle

In Ref.~\cite{Zheng05}, Zheng proposed a scheme for implementing a
conditional phase shift via adiabatic passages. The author claims
that the gate is ``neither of dynamical nor geometric origin'' on
the grounds that the Hamiltonian does not follow a cyclic change. He
further argues that ``in comparison with the adiabatic geometric
gates, the nontrivial cyclic loop is unnecessary, and thus the
errors in obtaining the required solid angle are avoided, which
makes this new kind of phase gates superior to the geometric
gates.'' In this Comment, we point out that geometric operations in
general, and adiabatic holonomies in particular, can be induced by
noncyclic Hamiltonians, and show that the gate in Zheng's scheme is
geometric. We also argue that the nontrivial loop responsible for
the phase shift is there, and it requires the same precision as in
any adiabatic geometric gate.

The scheme in \cite{Zheng05} involves two 4-level systems and one
qubit that have bases $\{|e_1\rangle,|e'_1\rangle,|g_1\rangle,
|g'_1\rangle \}$, $\{|e_2\rangle,|e'_2\rangle,|g_2\rangle,
|g'_2\rangle \}$, and $\{|0\rangle,|1\rangle\}$, respectively. Two
logical qubits are encoded in the subsystems with bases
$\{|e_10\rangle,|g_10\rangle\}$ and $\{|g_20\rangle, |g'_20\rangle
\}$. The Hamiltonian driving the evolution can be written
$H=\lambda_1|e_1 0\rangle\langle g_1 1|-\lambda_2|e_2
0\rangle\langle g_2 1|-\lambda_3 |e'_2 0\rangle\langle g'_2
1|+H.c.\notag$ The evolution has two stages. During the first stage,
the parameters $\lambda_1$, $\lambda_2$, $\lambda_3$ change
adiabatically so that $\theta=\arccos({\lambda_2/\sqrt{\lambda_1^2+
\lambda_2^2}})$ and $\theta'=\arccos({\lambda_3/\sqrt{\lambda_1^2+
\lambda_3^2}})$ change from $0$ to $\pi/2$. In the second stage, the
parameters change adiabatically so that $\theta$ changes from
$\pi/2$ to $0$, while $\theta'$ changes from $\pi/2$ to $\pi$. The
logical information is contained in the dark subspace
$\textrm{Span}\{|l_1\rangle, |l_2\rangle
,|l_3(\theta)\rangle,|l_4(\theta')\rangle\}$, where
$|l_1\rangle\equiv |g_1g_20\rangle$, $|l_2\rangle\equiv
|g_1g'_20\rangle$, $|l_3(\theta)\rangle\equiv
\cos{\theta}|e_1g_20\rangle+\sin{\theta}|g_1e_20\rangle$,
$|l_4(\theta')\rangle\equiv\cos{\theta'}|e_1g'_20\rangle+\sin{\theta'}|g_1e'_20\rangle$.
In the adiabatic limit, this subspace is decoupled from the rest of
the Hilbert space and its evolution results in the conditional phase
shift $|g_1g_20\rangle\rightarrow |g_1g_20\rangle$,
$|g_1g'_20\rangle\rightarrow |g_1g'_20\rangle$,
$|e_1g_20\rangle\rightarrow |e_1g_20\rangle$,
$|e_1g'_20\rangle\rightarrow -|e_1g'_20\rangle$. Since the states of
interest evolve in the dark space, no dynamical phases contribute to
the logical transformation. According to Zheng, ``since no solid
angle is swept in the parametric space, no Berry geometric phase
\cite{Berry84} is involved'' either.

Indeed, the Hamiltonian does not undergo a cyclic change. However,
Berry's phase has been extended to cyclic evolutions driven by not
necessarily cyclic Hamiltonians \cite{AharonovAnandan}, as well as
to noncyclic (both nonadiabatic \cite{Samuel} and adiabatic
\cite{Polavieja}) evolutions, and to the nonabelian generalizations
of these \cite{WZ,Mostafazadeh,Kult}. As shown below, the phase shift in Zheng's scheme is geometric whether looked upon as resulting from a path in the space of control parameters of the Hamiltonian, or a path in a Grassmannian.

First, the transformation in the logical space can be understood as
a holonomy resulting from parallel transport of vectors along an
\textit{open path} in the bundle defined by the eigenspace
$\textrm{Span}\{|l_1\rangle, |l_2\rangle
,|l_3(\theta)\rangle,|l_4(\theta')\rangle\}$ over the space of
parameters $\Lambda=\{(\lambda_1,\lambda_2,\lambda_3)\in R^3:
\sqrt{\lambda_1^2+\lambda_2^2}\neq 0,
\sqrt{\lambda_1^2+\lambda_3^2}\neq 0\}$. This picture can be
simplified since for all times the Hamiltonian has the
block-diagonal form
$H(t)=\textrm{diag}\{0,0,H_{1}(t),H_{2}(t),...\}$, where the first
two zeros correspond to $\textrm{Span}\{|l_1\rangle\}$ and
$\textrm{Span}\{|l_2\rangle\}$, $H_{1}(t)$ corresponds to
$\textrm{Span}\{|e_1g_20\rangle, |g_1e_20\rangle, |g_1g_21\rangle\}$
where the only dark state is $|l_3(\theta)\rangle$, and $H_{2}(t)$
corresponds to $\textrm{Span}\{|e_1g'_20\rangle, |g_1e'_20\rangle,
|g_1g'_21\rangle\}$ where the only dark state is
$|l_4(\theta')\rangle$. Thus the four logical states are decoupled
and it suffices to look at the geometric phases acquired by each of
them individually, which result from parallel transport in the
corresponding 1-dimensional bundles over $\Lambda$. The only
non-trivial loop occurs in the line bundle defined by
$\textrm{Span}\{|l_4(\theta')\rangle\}$. There, the initial state
$|l_4(0)\rangle$ is parallel-transported with the parallel condition
being $\langle
l_4(\theta'(s))|\frac{d}{ds}|l_4(\theta'(s))\rangle=0$, where $s\in
[0,1]$ parametrizes the path in $\Lambda$. Let us denote the basis
along the path by
$|{\psi}(s)\rangle=e^{i\phi(s)}|l_4(\theta'(s))\rangle$. The initial
and final points in ${\Lambda}$ are not the same but one can obtain
a gauge-invariant expression for the geometric phase associated with
the path by fixing the basis at the final point to be the one which
is `most parallel' to the initial basis, i.e., the one which
minimizes $\parallel\! |{\psi}(0)\rangle-|{\psi}(1)
\rangle\!\parallel$ \cite{Kult}. The geometric phase is then
${\beta}=\textrm{arg}\langle{\psi}(0)|{\psi}(1)
\rangle+i\int_0^1ds\langle{\psi}(s)|\frac{d}{ds}|{\psi}(s)\rangle$.
Since here the initial and final fibers are identical, the `most
parallel' choice for the initial and final frames is
$|{\psi}(0)\rangle=|{\psi}(1) \rangle$
($\textrm{arg}\langle{\psi}(0)|{\psi}(1) \rangle=0$). The expression
thus reduces to the Berry formula \cite{Berry84} which for this case
yields ${\beta}=\pi$.

Alternatively, the gate can be understood as a \textit{closed-loop}
holonomy in the tautological bundle over the Grassmannian
$\mathcal{G}(32,4)$ parametrizing the set of 4-dimensional subspaces
of the full Hilbert space $\mathcal{H}$. This picture emphasizes
that what is relevant for the holonomy in an adiabatically decoupled
eigenspace is how this subspace changes in $\mathcal{H}$
\cite{Kult}. Here, the logical space undergoes a closed loop (the
Hamiltonian is noncyclic only in a subspace which is adiabatically
decoupled from the logical subspace). That loop
requires the same precision as in any adiabatic geometric gate. In particular, the acquired geometric phase equals half of the solid angle enclosed by $|l_4(\theta')\rangle$ in the Bloch sphere (the projective Hilbert space
$\mathcal{G}(2,1)$) of $\textrm{span}\{|e_1g'_2 0\rangle, |g_1e'_20\rangle\}$.\\


\begin{thebibliography}{99}

\bibitem{Zheng05} S.-B. Zheng, Phys. Rev. Lett. \textbf{95}, 080502
(2005).

\bibitem{Berry84} M. Berry, Proc. R. Soc. Lond. A {\bf 392}, 45
(1984).

\bibitem{AharonovAnandan} Y. Aharonov and J. Anandan, Phys. Rev. Lett. \textbf{58}, 1593
(1987).

\bibitem{Samuel} J. Samuel, R. Bhandari, Phys. Rev. Lett. \textbf{60}, 2339 (1988).

\bibitem{Polavieja} G. G. de Polavieja and E. Sj¨oqvist, Am. J. Phys. \textbf{66}, 431 (1998).

\bibitem{WZ} F. Wilczek and A. Zee, Phys. Rev. Lett. \textbf{52}, 2111
(1984).

\bibitem{Mostafazadeh} A. Mostafazadeh, J. Phys. A \textbf{32}, 8157
(1999).

\bibitem{Kult} D. Kult, J. {\AA}berg, and E. Sj\"{o}qvist, Phys. Rev. A
\textbf{74}, 022106 (2006).





\end{thebibliography}
\end{document}